\documentclass{amsart}

\usepackage{mathptmx}        
\usepackage{helvet}          
\usepackage{courier}         

\usepackage{graphicx}        
\usepackage{multicol}        
\usepackage[bottom]{footmisc}
\usepackage{amsmath,amssymb}

\def\m{\bf m}
\def\e{\bf e}
\def\F{\bf F}
\def\C{\mathbf{C}}
\def\M{\bf M}
\def\I{\bf I}
\def\P{\bf P}

\newcommand{\tr}{\operatorname{tr}}

\newcommand{\de}[2]{\frac{\partial #1}{\partial #2}}

\title{A continuum model of skeletal muscle tissue with loss of activation}
\author{Giulia Giantesio \and Alessandro Musesti}
\address{Dipartimento di Matematica e Fisica, Universit\`a Cattolica
  del Sacro Cuore, via Musei 41, 25121 Brescia, Italy}

\email{giulia.giantesio@unicatt.it,alessandro.musesti@unicatt.it}

\begin{document}

\maketitle

\begin{abstract}We present a continuum model for the mechanical behavior of
  the skeletal muscle tissue when its functionality is reduced due to
  aging. The loss of ability of activating is typical of the geriatric
  syndrome called sarcopenia.  The material is described by a
  hyperelastic, polyconvex, transverse isotropic strain energy
  function. The three material parameters appearing in the energy are
  fitted by least square optimization on experimental
  data, while incompressibility is assumed through a Lagrange multiplier
  representing the hydrostatic pressure. The activation of the muscle
  fibers, which is related to the contraction of the sarcomere, is
  modeled by the so called active strain approach.  The loss of
  performance of an elder muscle is then obtained by lowering of some
  percentage the active part of the stress. The model is implemented
  numerically and the obtained results are discussed and graphically
  represented.
\end{abstract}

\section{Introduction}
\label{D:sec:intro}
Skeletal muscle tissue is one of the main components of the human
body, being about 40\% of its total mass. Its principal role is the
production of \emph{force}, which supports the body and becomes
\emph{movement} by acting on bones. The mechanism by which a muscle
produces force is called \emph{activation}.

Skeletal muscle tissue is a highly ordered hierarchical structure. The
cells of the tissue are the muscular fibers, having a length up to
several centimeters; they are organized in fascicles, where every
fiber is multiply connected to nerve axons, which drive the activation
of the tissue. Connective tissue, which is essentially isotropic,
fills the spaces among the fibers. Every fiber contains a
concatenation of millions of sarcomeres, which are the fundamental unit
of the muscle. With a length of some micrometers, a sarcomere is
composed by chains of proteins, mainly actin and myosin, which can
slide on each other. This sliding movement produces the contraction of
the sarcomere and, ultimately, the contraction of the whole muscle and
the production of force and movement.

The aim of this Chapter is to propose a mathematical model of skeletal
muscle tissue with a reduced activation, which is typical of a
geriatric syndrome named \emph{sarcopenia}\index{sarcopenia} \cite{D:giulio}.  About
thirty years ago, the term sarcopenia (from Greek \emph{sarx} or flesh and
\emph{penia}
or loss) has been introduced in order to describe the age-related
decrease of muscle mass and performance. Sarcopenia has since then
been defined as the loss of skeletal muscle mass and strength that
occurs with advancing age, which in turn affects balance, gait and
overall ability to perform even the simple tasks of daily living such
as rising from a chair or climbing steps. According to
\cite{D:reportSarcopenia}, sarcopenia affects more than 50 millions
people today and it will affect more than 200 millions in the next 40
years. There is still no generally accepted test for its diagnosis and
many efforts are made nowadays by the medical community to better
understand this syndrome. Therefore it is desirable to build a
mathematical model of muscle tissue affected by sarcopenia.  However,
to the best of our knowledge, in the biomathematical literature the
topic of loss of activation has never been addressed.

In order to use the valuable tools of Continuum Mechanics, during the
last decades the skeletal muscle tissue has been often modeled as a
continuum material \cite{D:ebi,D:thomas,D:bol,D:review}, 
which is usually assumed to be transversely
isotropic and incompressible. The former assumption is motivated by
the alignment of the muscular fibers, while the latter is ensured by
the high water content of the tissue (about 75\% of the total volume).
Moreover, in view of some experimental tests, the material is assumed
to be nonlinear and viscoelastic. Focusing our attention only on the
steady properties of the tissue, here we neglect the viscous effects
and we set in the framework of hyperelasticity.

In the model that we propose, there are three constitutive
prescriptions: one for the hyperelastic energy when the tissue is not
active (\emph{passive energy}), one for the activation and one for the loss
of performance. As far as the passive part is concerned, we assume an
exponential stress response of the material, which is customary in
biological tissues. The particular form that we choose, being a slight
simplification of the one proposed in~\cite{D:ebi}, has the 
advantage of being polyconvex and coercive, giving mathematical
soundness to the model and stability to the numerical
simulations.

A recent and very promising way to describe the activation is the
\emph{active strain} approach, where the extra energy produced by the
activation mechanism is encoded in a multiplicative decomposition of
the deformation gradient in an elastic and an active part (see
Section~\ref{D:subsec:activestrainmodel}). Unlike the classical
\emph{active stress} approach, in which the active part of the stress
is modeled in a pure phenomenological way and a new term has to be
added to the passive energy, the active strain method does not change
the form of the elastic energy, keeping in particular all its
mathematical properties. Moreover, at least in the case of skeletal
muscles, the active strain approach seems to be more adherent to the
physiology of the tissue, in the sense that at the molecular level the
production of force is actually given by a deformation of the
material, thanks to the contraction of the sarcomeres. The active part
of the deformation gradient is a mathematical representation of such a
contraction. The multiplicative decomposition of the deformation
gradient has been applied to an active striated muscle in
\cite{D:PTD,D:hernandez}. However, this decomposition involves only a
part of the whole elastic energy, which is written as the sum of two
terms for the case of a fiber-reinforced material. As far as we know,
the active strain approach has never been previously applied to the
whole elastic energy of a skeletal muscle tissue.  As a drawback, the
active strain approach can be a source of some technical difficulties;
for instance, in our case fitting the model on the experimental data
is not so simple, see eq.~\eqref{D:eqgammaesplicita}.

Furthermore, we consider the loss of performance, which is one of the
novelties of our model. Unfortunately, there are no experimental data
on the elastic properties of a sarcopenic muscle tissue, at least to
our knowledge; hence we adopt the naive strategy of reducing the
active part of the stress (which is the difference between the stress
of the material with and without activation) by a given percentage, represented
by the damage parameter $d$ (see
Section~\ref{D:subsec:lossactivation}). In this way, there is a single
parameter in the model which concisely accounts for any effect of the
disease.

The proposed model can be numerically implemented using finite element
methods. In Section~\ref{D:sec:numerical} we present some results
obtained using FEniCS, an open source collection of Python
libraries. Actually, we consider a cylindrical geometry with radial
symmetry, so that the numerical domain is two-dimensional and the
computational cost is reduced. As far as the boundary conditions are
concerned, we prescribe the displacement on the bases of the cylinder
and let the lateral surface traction-free. Such simulations show that
the experimental results of~\cite{D:datib} on the passive and active
stress-strain healthy curves, obtained \emph{in vivo} from a tetanized
tibialis anterior of a rat, can be well reproduced by our
model. Further, the behavior of the tissue when $d$ increases is
analyzed. An ongoing task is to perform a finite element
implementation of the model when generic loads are applied, and to
consider a realistic three-dimensional muscle mesh.  We are now
developing a truly hyperelastic model, where the expression of the
stress takes into account also the dependence of the activation on the
deformation gradient. Actually, in this chapter the stress is computed
as the derivative of the hyperelastic energy keeping the active part
of the deformation gradient fixed.

In the future, it will be very interesting to find some connections
between the damage percentage (the parameter $d$) and other
physiological quantities, such as the mass of the muscular tissue or
the neuronal activity. Another important topic will be the application
of some homogenization techniques in order to deduce an improved
constitutive equation for the skeletal muscle starting from its
microstructure.

\section{Constitutive model}
\label{D:sec:1}
Skeletal muscle tissue is characterized by densely packed muscle
fibers,\index{muscle fibers} which are arranged in fascicles. Filling the spaces between
the fibers and fascicles, connective tissue surrounds the muscle and
it is responsible of the elastic recoil of the muscle to elongation.
Besides a large amount of water, the fibers themselves contain titin,
actin and myosin filaments. The latter two sliding elements form the
actual contractile component of the muscle, which is called
\emph{sarcomere}. Since the fibers locally follow a predominant
unidirectional alignment, transverse isotropy with respect to that
main direction can be assumed. We hence begin by modelling the
skeletal muscle tissue as a transversely isotropic nonlinear
hyperelastic material with principal direction $\m$, which follows
the alignment of the muscle fibers.

\subsection{Passive model}
\label{D:subsec:passive}
\begin{figure}[htbp]
        \includegraphics[width=0.70\textwidth]{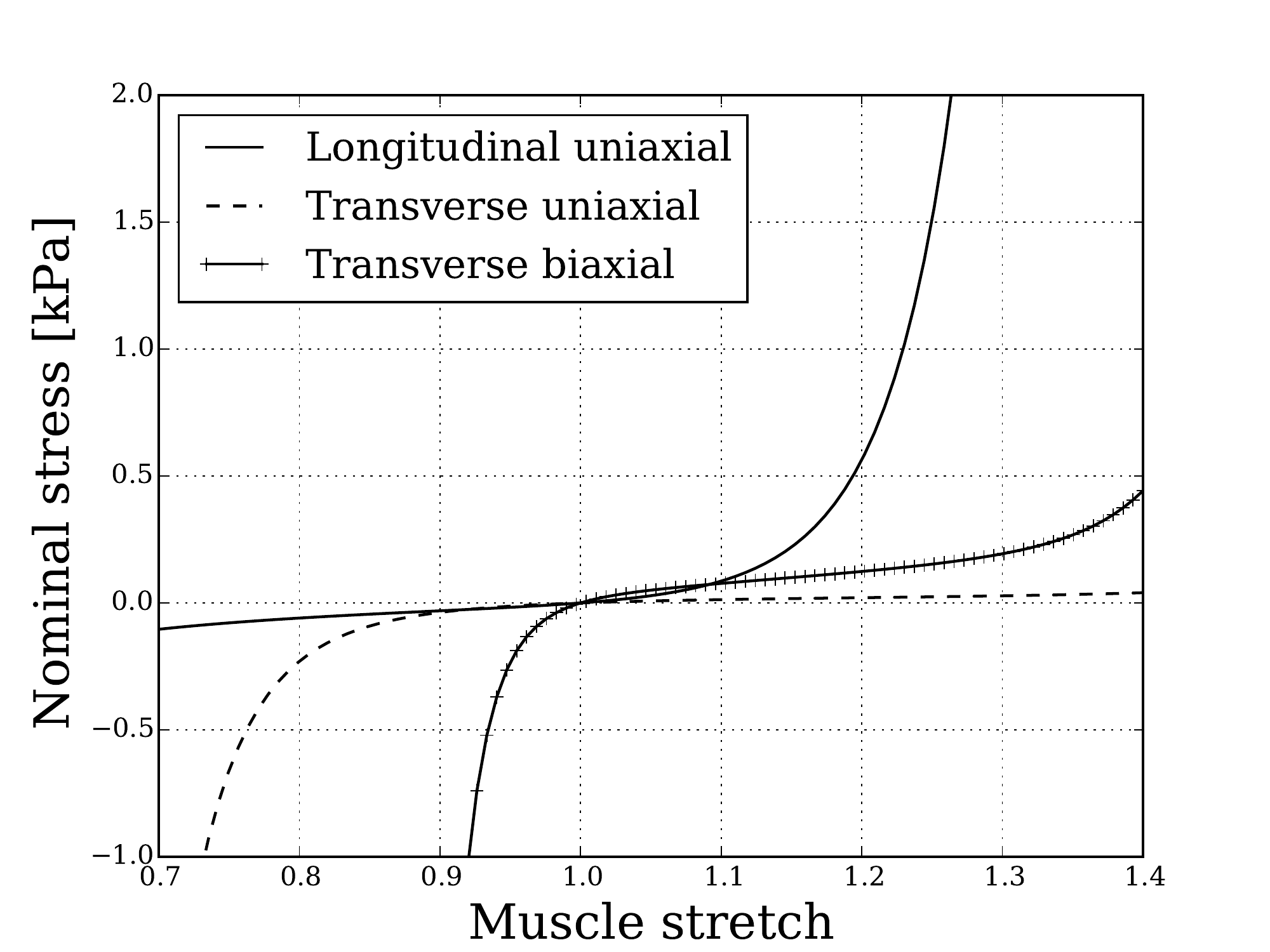}
   \caption{Transversely isotropic behavior of the model.}
     \label{D:biaxial}
\end{figure}

\index{passive energy}
Let $\F$ denote the deformation gradient tensor, $\C=\F^T\F$ the right
Cauchy-Green tensor and $\M=\m\otimes\m$ the so called {\em structural
  tensor}.\index{structural tensor}  If $\Omega$ denotes the reference configuration occupied
by the muscle, we describe its passive behavior by choosing a
hyperelastic strain energy function
\begin{equation}
 \int_{\Omega} W({\C})dV, \nonumber
\end{equation}
where the strain energy density is of the form 
\begin{equation}
W({\C})=\frac{\mu}{4}\left\{\frac{1}{\alpha}\left[e^{\alpha(I_p-1)}-1\right]+K_p-1\right\}, \label{D:Wpassiva}
\end{equation}
with
\begin{equation}
I_p=\frac{w_0}{3}\tr({\C})+(1-w_0)\tr({\C}{\M}), \ \
K_p=\frac{w_0}{3}\tr({{\C}^{-1}})+(1-w_0)\tr({\C}^{-1}\M). \nonumber
\end{equation}
Here $\mu$ is an elastic parameter and $\alpha$ and $w_0$ are positive
dimensionless material parameters. The generalized invariants $I_p$
and $K_p$ are given by a weighted combination of the isotropic and
anisotropic components; in particular, $w_0$ measures the ratio of
isotropic tissue constituents and $1-w_0$ that of muscle
fibers. Moreover, the term $\tr({\C\M})$ represents the squared
stretch in the direction of the muscle fiber and is thus associated with
longitudinal fiber properties, while the term $\tr({{\C}^{-1}\M})$
describes the change of the squared cross-sectional area of a surface
element which is normal to the direction $\m$ in the
reference configuration and thus relates to the transverse
behavior of the material \cite{D:sch,D:ei2009} (see Fig.~\ref{D:biaxial}).  

One of the mathematical features of the energy density
\eqref{D:Wpassiva} is that it is polyconvex and coercive
\cite{D:ebipcb, D:sch}, hence the equilibrium problem with mixed boundary
conditions is well posed.

We remark that ${\C}$ is the identity tensor ${\I}$ in the reference
configuration, so that $I_p=K_p=1$, \emph{i.e.} we have the energy- and
stress-free state of the passive muscle tissue (see \cite{D:ebipcb}).

The high content of water is responsible of the nearly incompressible
behavior which is experimentally reported for muscle fibers, so that
we can assume\index{incompressibility}
\begin{equation}
 \det {\C} = 1. \label{D:incomp} 
\end{equation}

As is customary in hyperelasticity, the first
Piola-Kirchhoff\index{Piola-Kirchhoff stress tensor} stress
tensor, known as \emph{nominal stress tensor}, can be directly
computed by differentiating the strain energy function:
\begin{eqnarray}
 {\P}
 &=&\de{W}{{\F}}-p{\F}^{-T}=2{\F}\de{W}{{\C}}-p{\F}^{-T}= \label{D:stressPK} \\[1ex]
 &=&\frac{\mu}{2}{\F}\left\{e^{\alpha(I_p-1)}\left[\frac{w_0}{3}{\I}+(1-w_0){\M}\right]-{\C}^{-1}\left[\frac{w_0}{3}{\I}+(1-w_0){\M}\right]{\C}^{-1}\right\}-p{\F}^{-T}, \nonumber
\end{eqnarray}
where $p$ is a Lagrange multiplier associated with the hydrostatic
pressure which results from the incompressibility constraint
\eqref{D:incomp}.

\begin{figure}[htbp]
        \includegraphics[width=0.70\textwidth]{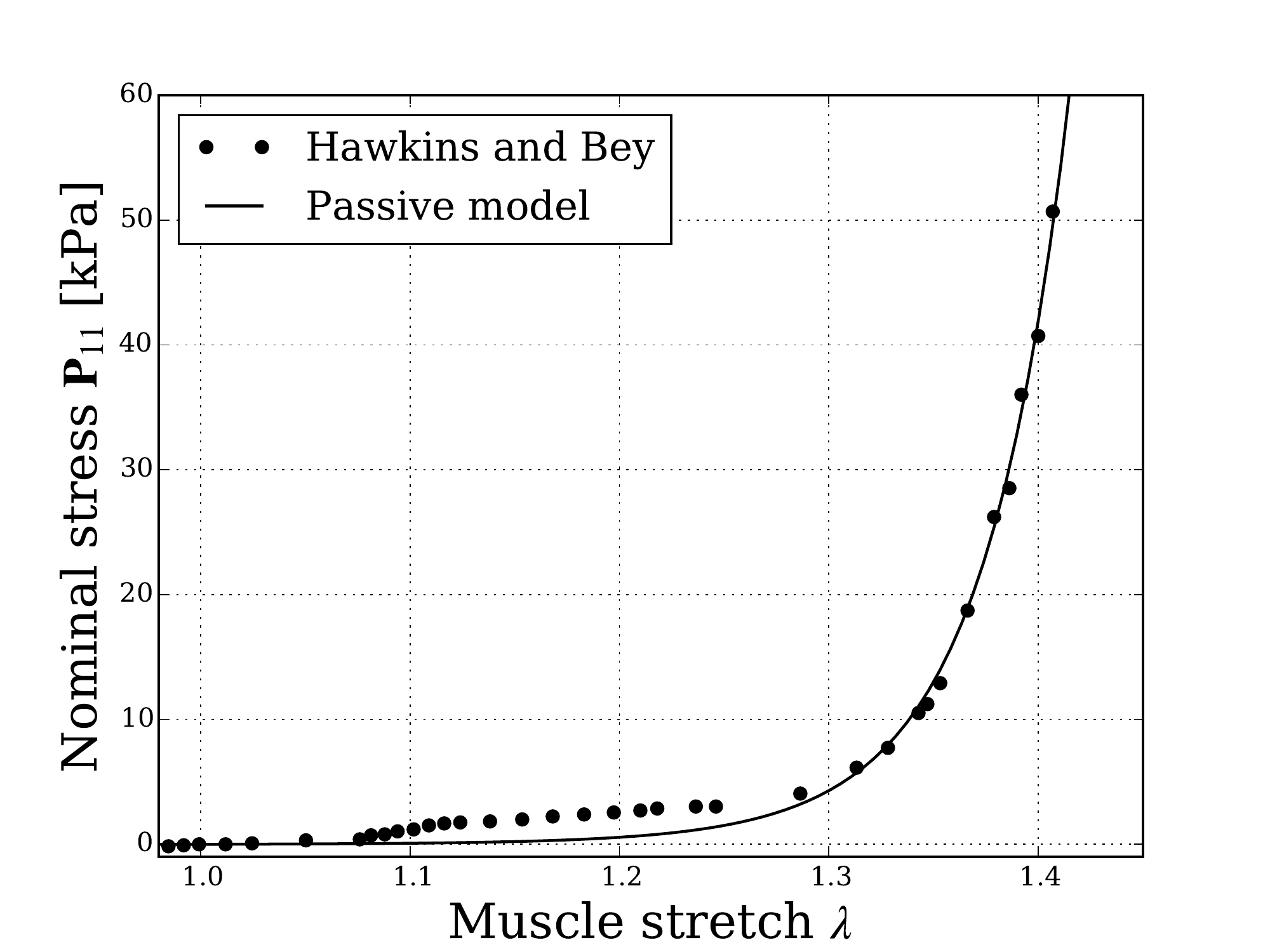}
   \caption{Comparison of the passive model in uniaxial tension with the experimental data of a rat tibialis anterior muscle reported in \cite{D:datib}.} 
     \label{D:passivedata}
\end{figure}

The material parameters of the model can be obtained from real data.
More precisely, concerning the elastic parameter $\mu$, we use the
value given in \cite{D:ebi}, while the other two parameters have been
obtained by least squares optimization using the experimental data by
Hawkins and Bey \cite{D:datib} about the stretch response of a
tetanized {\em tibialis anterior} of a rat (see
Fig.~\ref{D:passivedata}). In Table \ref{D:parpass} we furnish the
values of the parameters.

\begin{table}
\label{D:parpass}
\begin{tabular}{p{3.75cm}p{3.75cm}p{3.75cm}}
\hline\noalign{\smallskip}
$\mu$ [kPa] & $\alpha$ [-] & $w_0$ [-]\\
\noalign{\smallskip}\hline\noalign{\smallskip}
0.1599 & 19.35 & 0.7335 \\
\noalign{\smallskip}\hline\noalign{\smallskip}
\end{tabular}
\caption{Material parameters of the passive model.}
\end{table}

\begin{figure}[htbp]
        \includegraphics[width=0.70\textwidth]{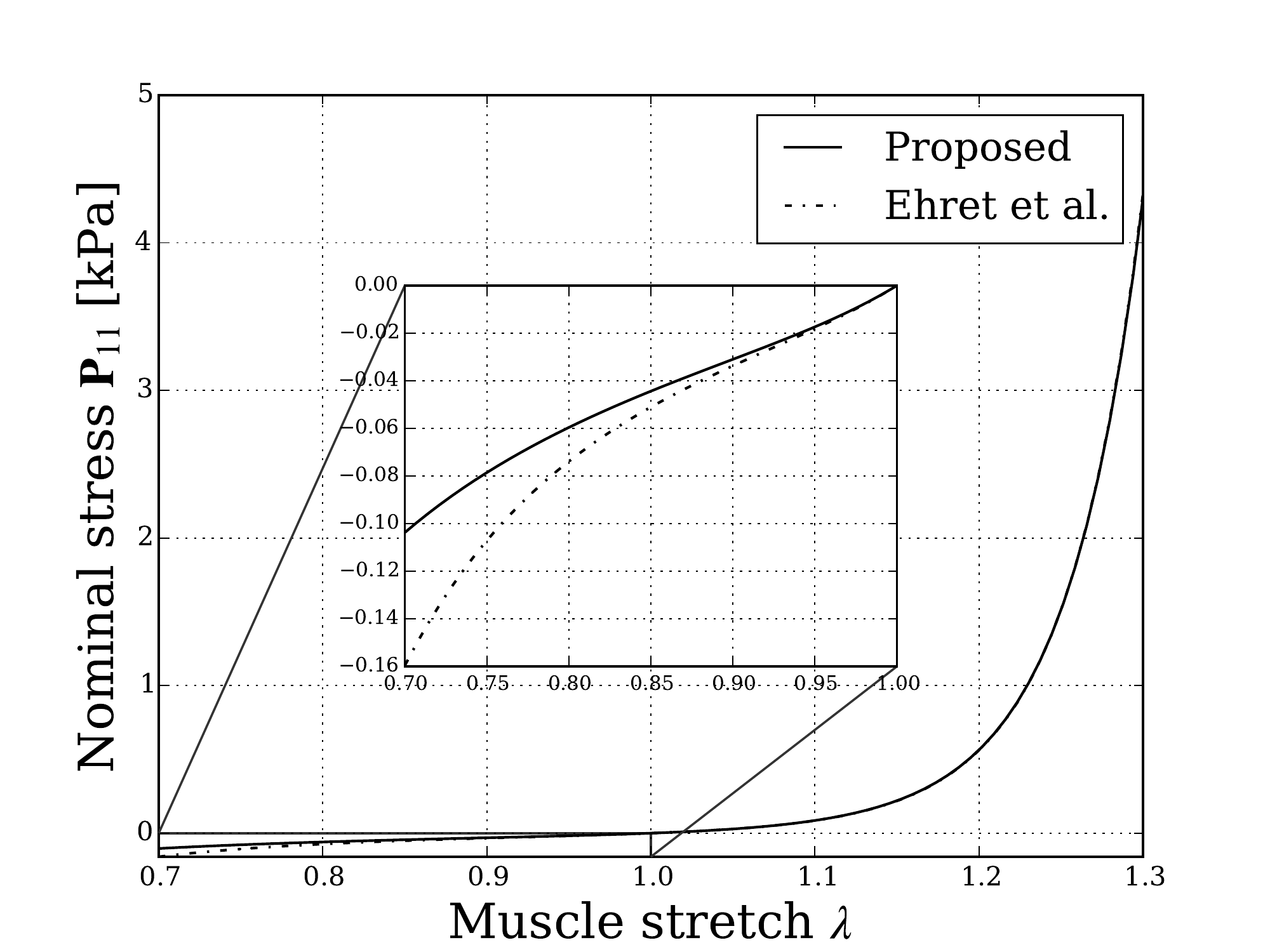}
   \caption{Comparison between the passive stress here proposed and the one studied in \cite{D:ebi} during uniaxial tension along the fibers.} 
     \label{D:paragoneebi}
\end{figure}

We remark that the strain energy function \eqref{D:Wpassiva} is a
slight simplification of the one proposed by Ehret, B\"ol and Itskov in \cite{D:ebi}:
\begin{equation}
W_\text{EBI}(\C)=\frac{\mu}{4}\left\{\frac{1}{\alpha}\left[e^{\alpha(I_p-1)}-1\right]+
\frac{1}{\beta}\left[e^{\beta(K_p-1)}-1\right]\right\}, \label{D:Webi}
\end{equation}
where $\alpha=19.69, \beta=1.190, w_0=0.7388$.  Actually, our
simplification consists in linearizing the term related to $K_p$,
which describes the transverse behavior. This is motivated by the
fact that the parameter $\beta$ is much smaller than $\alpha$. In
Fig.~\ref{D:paragoneebi} we can see the comparison between the nominal
stress in the direction of the stretch of the two models when the
muscle fibers are elongated in their direction.

\subsection{Active model}\label{D:subsec:activestrainmodel}
\index{activation}One of the main features of the skeletal muscle tissue is its ability
of being voluntarily activated. Skeletal muscles are activated
through electrical impulses from motor nerves; the activation triggers
a chemical reaction between the actin and myosin filaments which
produces a sliding of the molecular chains, causing a contraction of
the muscle fibers.

During the last decades, many authors tried to mathematically model
the process of activation, mainly with two different approaches (for a
review see \cite{D:asasb}).  The most famous approach followed in the
literature is called \emph{active stress} and it consists in adding an
extra term to the stress, which accounts for the contribution
given by the activation (see for example
\cite{D:martins,D:blemker,D:thomas}).  However, this is an {\em ad
  hoc} method, usually not related to the sliding movement of the
filaments in the sarcomeres, which is the main mechanism of
contraction at the mesoscale.

More recently, the \emph{active strain}\index{activation!active strain} 
approach was proposed by Taber
and Perucchio \cite{D:taber} in order to describe the activation of
the cardiac tissue, following previous theories of growth and
morphogenesis, as well as several models of plasticity. The method for
soft living tissues is explained in \cite{D:nardinocchiteresi}.
Differently from the active stress approach, this method does not
change the form of the strain energy function; rather, it assumes that
only a part of the deformation gradient, obtained by a multiplicative
decomposition, is responsible for the store of elastic energy.  This
method is related to the biological meaning of activation and can be
reasonably adopted also in our case. To the best of our knowledge, the
active strain approach has never been followed for the skeletal muscle
tissue in literature.

\begin{figure}[htbp]
        \includegraphics[width=0.70\textwidth]{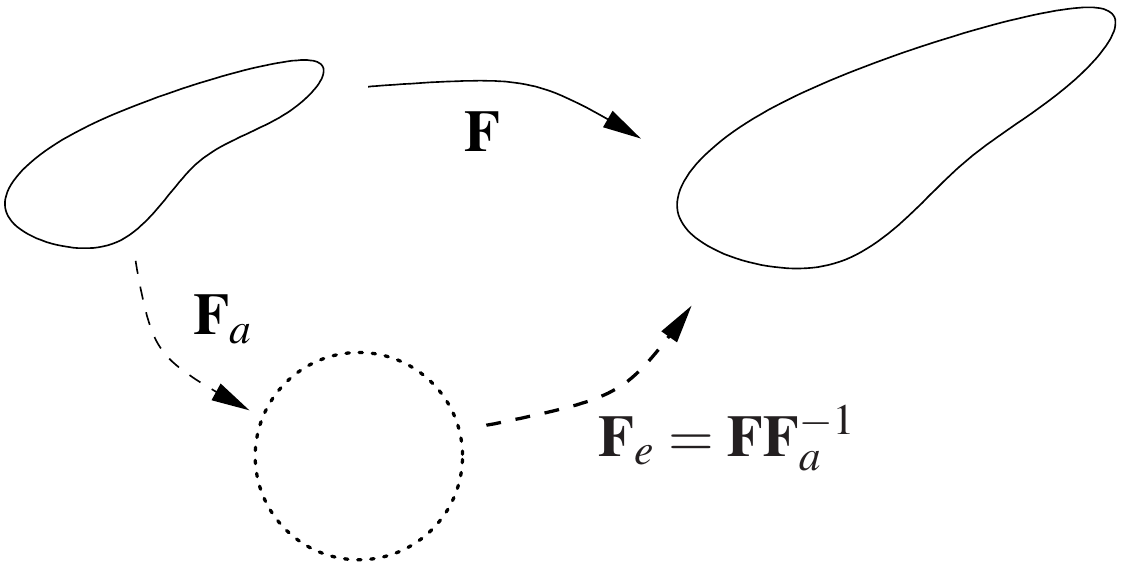}
   \caption{Pictorial view of the active strain approach.} 
     \label{D:figactivestrainapproach}
\end{figure}

We begin by rewriting the deformation gradient as ${\F}={\F}_e{\F}_a$, where
${\F}_e$ is the elastic part and ${\F}_a$ describes the active
contribution (see Fig.~\ref{D:figactivestrainapproach}). 
The active strain ${\F}_a$ represents a change of the reference
volume elements due to the contraction of the sarcomeres, so that it
does not contribute to the elastic energy.  
A reference volume element, distorted by ${\F}_a$, needs a further
deformation ${\F}_e$ to match the actual volume element,
which accommodates
both the external forces and the active contraction. Notice that neither ${\F}_a$ nor
${\F}_e$ need to be the gradients of some displacement, that is, it is
not necessary that they fulfill the compatibility condition
$\operatorname{curl}{\F}_a=0$ or $\operatorname{curl}{\F}_e=0$.

The volume elements are modified by the internal active forces without
changing the elastic energy, hence the strain energy function of the
activated material has to be computed using ${\C}_e={\F}_e^T{\F}_e$ and
taking into account ${\F}_e={\F}{\F}_a^{-1}$. If ${\F}_a=\mbox{grad}\chi_a$
for some displacement $\chi_a$, then from
Fig.~\ref{D:figactivestrainapproach} by a change of variables it is easy to see that
\begin{align}
\int_{\chi_a(\Omega)}W({\C}_e)d\widehat{V}=\int_{\Omega}W({\F}^{-T}_a{\C}{\F}_a^{-1})(\det{\F}_a)d{V}. \nonumber
\end{align}
The right-hand side of the previous equation is well defined also when
${\F}_a$ does not come from a global displacement, and it describes
the strain energy of the active body.
We then obtain the modified hyperelastic energy density  
\[
\widehat{W}({\C}) = (\det{\F}_a)W({\C}_e) = 
(\det{\F}_a)W({\F}^{-T}_a{\C}{\F}_a^{-1}).
\]

We now have to model the active part ${\F}_a$. 
Since the activation of the muscle consists in a contraction along the fibers, we choose
\begin{equation}
 {\F}_a={\I}-\gamma {\m}\otimes{\m}, \label{D:Fa} 
\end{equation}
where $0\leq\gamma <1$  is a dimensionless parameter representing
the relative contraction of activated fibers ($\gamma=0$ meaning no activation).
Then the modified strain energy density becomes 
\begin{eqnarray}
&&\widehat{W}({\C})=(1-\gamma)W({\C}_e)=(1-\gamma)\frac{\mu}{4}\left\{\frac{1}{\alpha}\left[e^{\alpha(I_e-1)}-1\right]+K_e-1
\right\}, \label{D:Wattiva}\\[1ex]
&&I_e=\frac{w_0}{3}\tr({\C}_e)+(1-w_0)\tr({\C}_e{\M}), \ \ 
K_e=\frac{w_0}{3}\tr({{{\C}_e}^{-1}})+(1-w_0)\tr({{\C}_e}^{-1}{\M}). \nonumber
\end{eqnarray}
The corresponding first Piola-Kirchhoff stress tensor\index{Piola-Kirchhoff stress tensor} is given by
\begin{eqnarray}
 \widehat{\P}
 =&&\det{\F}_a\de{W}{{\F}_e}{\F}_a^{-1}-\widehat{p}{\F}^{-T}= \label{D:stressPKtotale} \\[1ex]
 =&&\frac{\mu}{2}(1-\gamma){\F}_e\left\{e^{\alpha(I_e-1)}\left[\frac{w_0}{3}{\I}+(1-w_0){\M}\right]-{\C}_e^{-1}\left[\frac{w_0}{3}{\I}+(1-w_0){\M}\right]
 {\C}_e^{-1}\right\}{\F}_a^{-1}  \nonumber\\
 &&-\widehat{p}{\F}^{-T},\nonumber
\end{eqnarray}
where $\widehat{p}$ accounts for the incompressibility constraint
$\det\C=1$. Notice that, since the activation~\eqref{D:Fa} does not
preserve volume and the material has to be globally incompressible,
one has that $\det{\C}_e\neq 1$, so that the material is elastically
compressible. As far as the strain energy density is concerned, a
factor $(1-\gamma)$ appears in~\eqref{D:Wattiva} which keeps into
account the compressibility of ${\F}_a$. It would be interesting to
study also other kinds of passive energies, involving the quantity
$\det \C$, in order to better describe the elastic compressibility of
the material.

\begin{figure}[htbp]
        \includegraphics[width=0.70\textwidth]{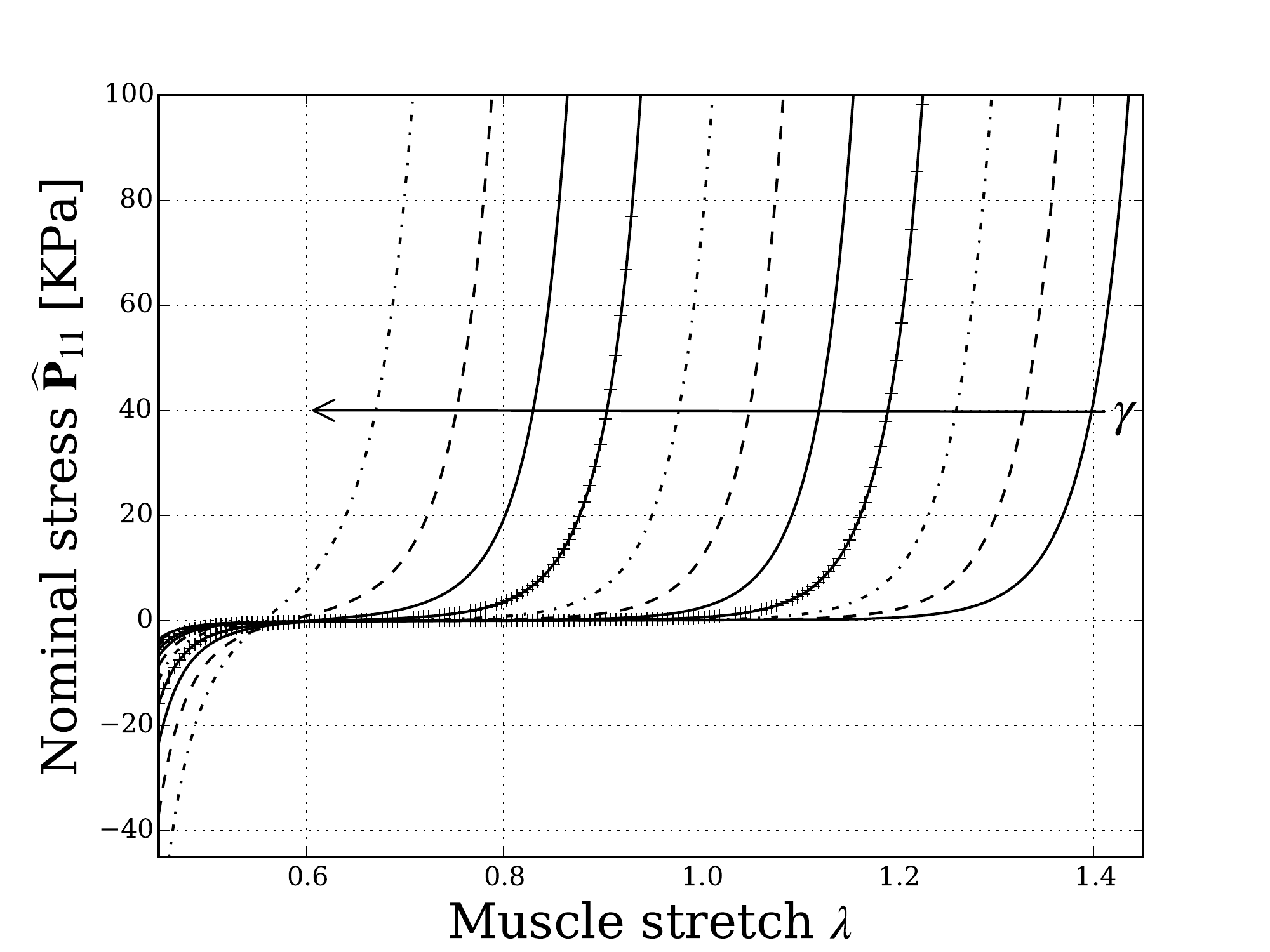}
   \caption{Stress-stretch curves in uniaxial tension for several values of $\gamma$ from $0$ to $0.4$.} 
     \label{D:gamma-s-s}
\end{figure}

In Fig.~\ref{D:gamma-s-s} we represent, for several values of the parameter
$\gamma$, the stress-strain curve for a uniaxial tension
along the fibers. If the muscle is activated ($\gamma >0$), then (the
absolute value of) the stress increases with $\gamma$ and the value of
the stretch such that the stress is zero becomes less than one.

\section{Modelling the activation on experimental data}
\label{D:sec:modelactivation}\index{activation}

\begin{figure}[htbp]
        \includegraphics[width=0.80\textwidth]{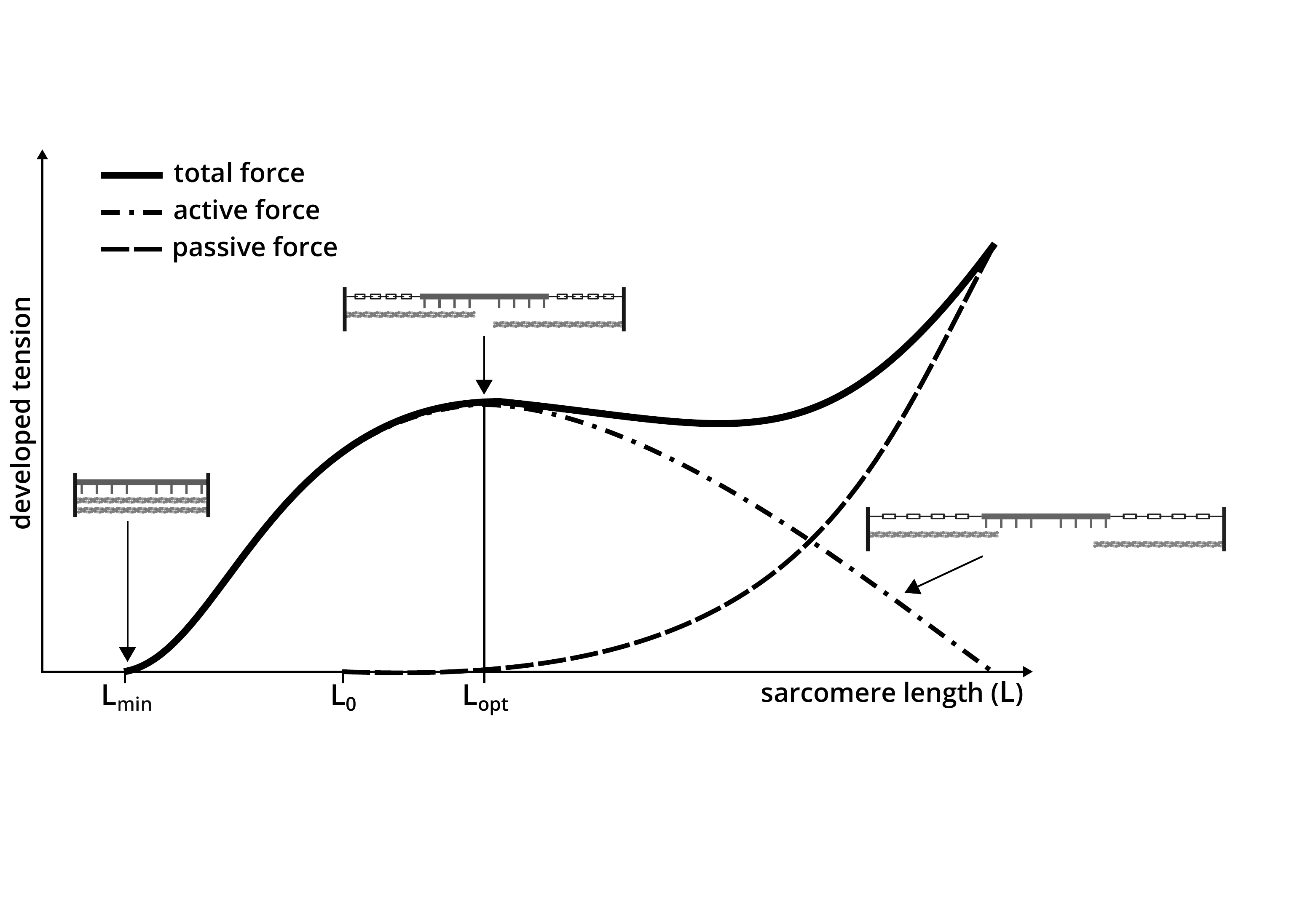}
        \caption{Length-tension relationship of a sarcomere. Here we
          denote by $\mathsf{L_{opt}}$ the length at which the
          sarcomere produces the maximum force in isometric
          experiments, by $\mathsf{L_0}$ the rest length and
          by $\mathsf{L_{min}}$ the minimal length of the sarcomere 
          (fully activated).}
     \label{D:attivovspassivo}
\end{figure}

The activation parameter $\gamma$, which was assumed constant in the
previous section, in fact usually depends on the deformation
gradient. In typical experiments on a tetanized skeletal muscle it is
apparent that the contraction of the fibers due to activation varies
with their stretch, reaching a maximum value and then
decreasing. Fig.~\ref{D:attivovspassivo} shows the qualitative
relation between the elongation and the developed stress.
This section will be devoted to taking into account this
phenomenon. Specifically, the expression of $\gamma$ will be
determined matching an experiment-based relation between stress and
strain with our model~\eqref{D:stressPKtotale}.

In order to find the relation between stress and strain, the
experiments {\em in vivo} are usually performed in two steps. First,
the stress-strain curve is obtained without any activation
(\emph{passive curve}). Second, by an electrical stimulus the muscle
is isometrically kept in a tetanized state and the \emph{total
  stress-strain curve} is plotted. The last curve, which is
qualitatively represented in Fig.~\ref{D:attivovspassivo}, depends on
the reciprocal position of actin and myosin chains. By taking the
difference of the two curves one can obtain the \emph{active curve},
describing the amount of stress due to activation. It is useful to
find a mathematical expression of such a curve, in order to take into
account the experimental behavior of the active contraction. This
issue has already been addressed in several papers, see
e.g. \cite{D:ebi,D:thomas,D:vanLeeuwen1991,D:vanLeeuwen1992,D:Johansson,D:blemker,D:bol}.

\begin{figure}[htbp]
        \includegraphics[width=0.70\textwidth]{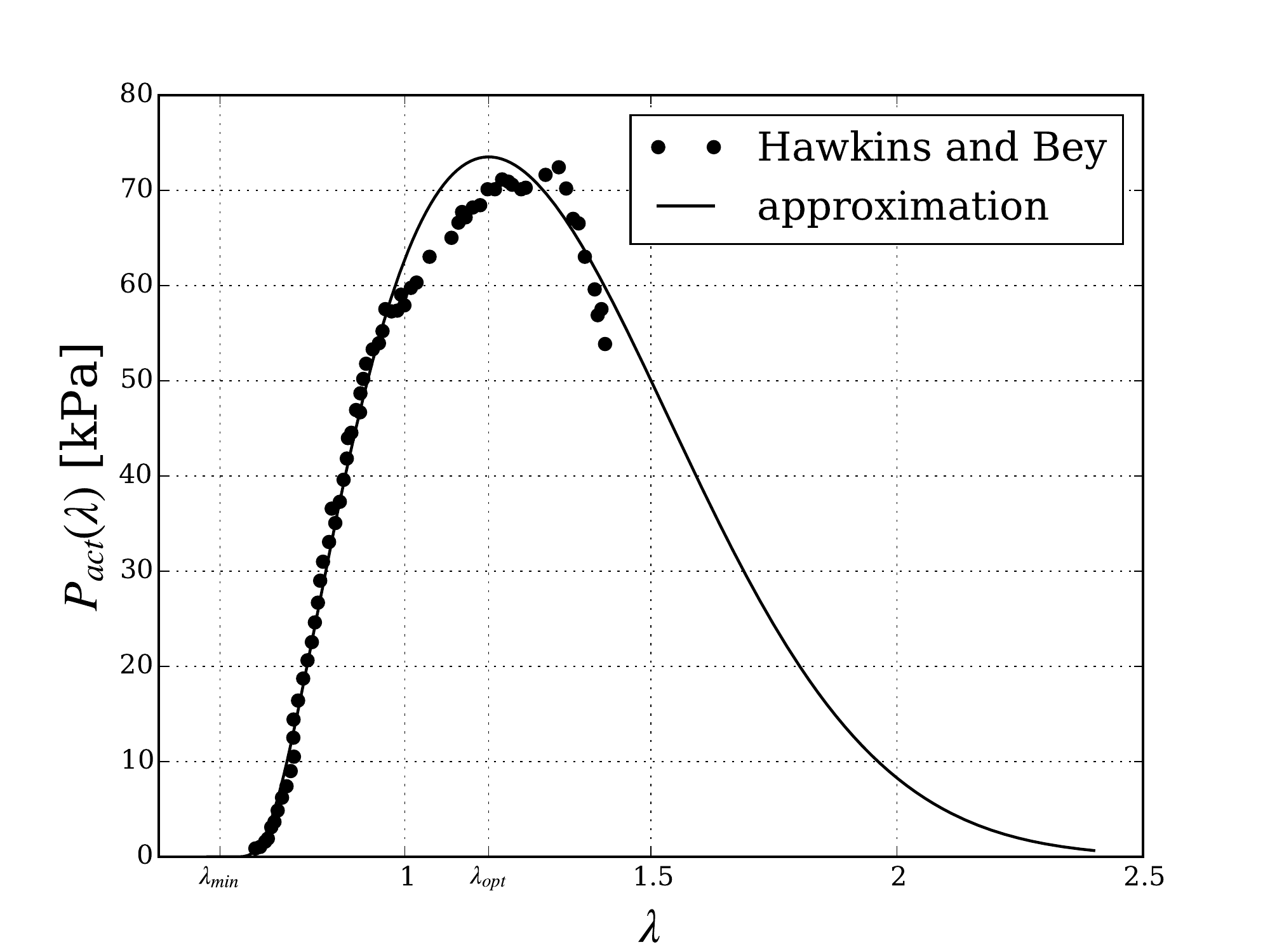}
   \caption{Plot of the active curve \eqref{D:pact} with the parameters reported in Table \ref{D:paratt} together with the representation 
   of the experimental data given in \cite{D:datib}.} 
     \label{D:fattiva}
\end{figure}

Denoting with $\lambda$ the ratio between the current length of the
muscle and its original length, we assume the active curve to be of
the form
\begin{equation}\label{D:pact}
 P_{act}(\lambda) =
\left\{
\begin{aligned}
&P_{opt}\exp\left[-k\frac{(\lambda^2-\lambda_{opt}^2)^2}{\lambda-\lambda_{min}}\right]
& \text{if $\lambda>\lambda_{min}$},
\\[1ex]
&0 & \text{otherwise},
\end{aligned}
\right. 
\end{equation}
where $\lambda_{min}$ is the minimum stretch value after which the
activation starts (\emph{i.e.}  the lower bound for the stretch at which the
myofilaments begin to overlap) and $k$ is merely a fitting
parameter. The coordinates $(\lambda_{opt},P_{opt})$ identify the
position of the maximum of the curve.  As it is explained in
\cite{D:ebi}, the value of $P_{opt}$ takes into account some
information at the mesoscale level, such as the number of activated
motor units and the interstimulus interval; according to the
literature \cite{D:ebi,D:thomas}, it is set at $P_{opt}=73.52$
kPa. The numerical values of the other three parameters, deduced
through least squares optimization on the data reported in
\cite{D:datib}, are given in Table \ref{D:paratt}.
\begin{table}
\caption{Material parameters of the active model.}
\label{D:paratt}
\begin{tabular}{p{2.82cm}p{2.82cm}p{2.82cm}p{2.82cm}}
\hline\noalign{\smallskip}
$\lambda_{min}$ [-] & $\lambda_{opt}$ [-] & $k$ [-] & $P_{opt}$ [kPa]\\
\noalign{\smallskip}\hline\noalign{\smallskip}
0.6243 & 1.1704 & 0.4342 &73.52 \\
\noalign{\smallskip}\hline\noalign{\smallskip}
\end{tabular}
\end{table}
The expression \eqref{D:pact} has the advantage of describing the
asymmetry between the ascending and descending branches of the active
curve obtained in \cite{D:datib}. Indeed, even if the asymmetry is not
so evident in their curve, due to the fact that there are only few
data on the descending branch, it is a typical feature of several
experimentally measured sarcomere length-force relation.  Moreover, as
one can easily see in Fig.~\ref{D:fattiva}, the convex behavior of the
data nearby $\lambda_{min}$ is well fitted.

\subsection{The activation parameter $\gamma$ as a function of the elongation}
\label{D:subsec:gammavselongation}\index{activation!parameter of}

Now our aim is to obtain $P_{act}(\lambda)$ given in \eqref{D:pact}
from the model described in Section
\ref{D:subsec:activestrainmodel}. In order to reach our purpose, we
have to model the activation parameter $\gamma$ as a function of the
stretch.

As in the experiments of Hawkins and Bey \cite{D:datib}, let us consider a uniaxial simple tension along the fibers. For simplicity, we assume that the fibers follow the direction 
${\m}={\e}_1$. 
Since the skeletal muscle tissue is modeled as an incompressible transversely isotropic material, the general form of the 
deformation gradient ${\F}$ is given by  
\[
{\F}=
\begin{pmatrix}
\lambda & 0 & 0 \\
0 & \frac{1}{\sqrt{\lambda}} & 0 \\
0 & 0 & \frac{1}{\sqrt\lambda}
\end{pmatrix}.
\]

Then using the notation introduced in Section \ref{D:subsec:activestrainmodel}, one has
\begin{eqnarray}
&{\C}_e&=
\begin{pmatrix}
\frac{\lambda^2}{(1-\gamma)^2} & 0 & 0 \\
0 & \frac{1}{\lambda} & 0 \\
0 & 0 & \frac{1}{\lambda}
\end{pmatrix}, \nonumber \\[1ex]
&I_e&=\frac{w_0}{3}\left[\frac{\lambda^2}{(1-\gamma)^2}+\frac{2}{\lambda}\right]+(1-w_0)\frac{\lambda^2}{(1-\gamma)^2}, \nonumber \\[1ex]
&K_e&=\frac{w_0}{3}\left[\frac{(1-\gamma)^2}{\lambda^2}+2\lambda\right]+(1-w_0)\frac{(1-\gamma)^2}{\lambda^2}. \nonumber
\end{eqnarray}

In this case, it is convenient to look at the strain energy as a function of the stretch $\lambda$ and the activation parameter $\gamma$:
\begin{equation}
\widehat{W}(\lambda,\gamma)=(1-\gamma)W(\lambda,\gamma)=(1-\gamma)\frac{\mu}{4}\left\{\frac{1}{\alpha}\left[e^{\alpha(I_e-1)}-1\right]+K_e-1
\right\}. \label{D:Wattgamma}
\end{equation}
 
Then the nominal stress along the fiber direction is given by
\begin{equation}
 P_{tot}(\lambda,\gamma):= \de{\widehat{W}}{\lambda}
 =(1-\gamma)\frac{\mu}{4}\left[I_e'
e^{\alpha(I_e-1)} +K_e'\right], \label{D:Ptotcalcolato}
\end{equation}
where 
\begin{eqnarray*}
&I_e'&=\de{I_e}{\lambda}=2\frac{w_0}{3}\left[\frac{\lambda}{(1-\gamma)^2}-\frac{1}
    {\lambda^2}\right]+2(1-w_0)\frac{\lambda}{(1-\gamma)^2},\\
&K_e'&=\de{K_e}{\lambda}=2\frac{w_0}{3}\left[-\frac{(1-\gamma)^2}{\lambda^3}+1\right]
-2(1-w_0)\frac{(1-\gamma)^2}{\lambda^3}.
\end{eqnarray*}

We can get the passive stress by setting $\gamma=0$: 
\begin{eqnarray}
P_{pas}(\lambda):=P_{tot}(\lambda,0)
=&\dfrac{\mu}{2}&\left\{\left[\left(1-\frac{2}{3}w_0\right)\lambda-\frac{w_0}{3}\frac{1}{\lambda^2}
\right]e^{\alpha\left[\left(1-\frac{2}{3}w_0\right)\lambda^2+\frac{w_0}{3}\frac{2}{\lambda}-1\right]}\right.\nonumber\\
&-&\left.\left(1-\frac{2}{3}w_0\right)\frac{1}{\lambda^3}+\frac{w_0}{3}\right\}.
\label{D:Ppas}
\end{eqnarray}
We remark that the values of $P_{tot}$ and $P_{pas}$ can also be
obtained by computing the first component of the stress given by
\eqref{D:stressPKtotale} and \eqref{D:stressPK} after finding the
hydrostatic pressure from the conditions
$\widehat{P}_{22}=\widehat{P}_{33}={P}_{22}={P}_{33}=0$ (traction-free
lateral surface).

\begin{figure}[htbp]
        \begin{tabular}{cc}
        \includegraphics[width=0.70\textwidth]{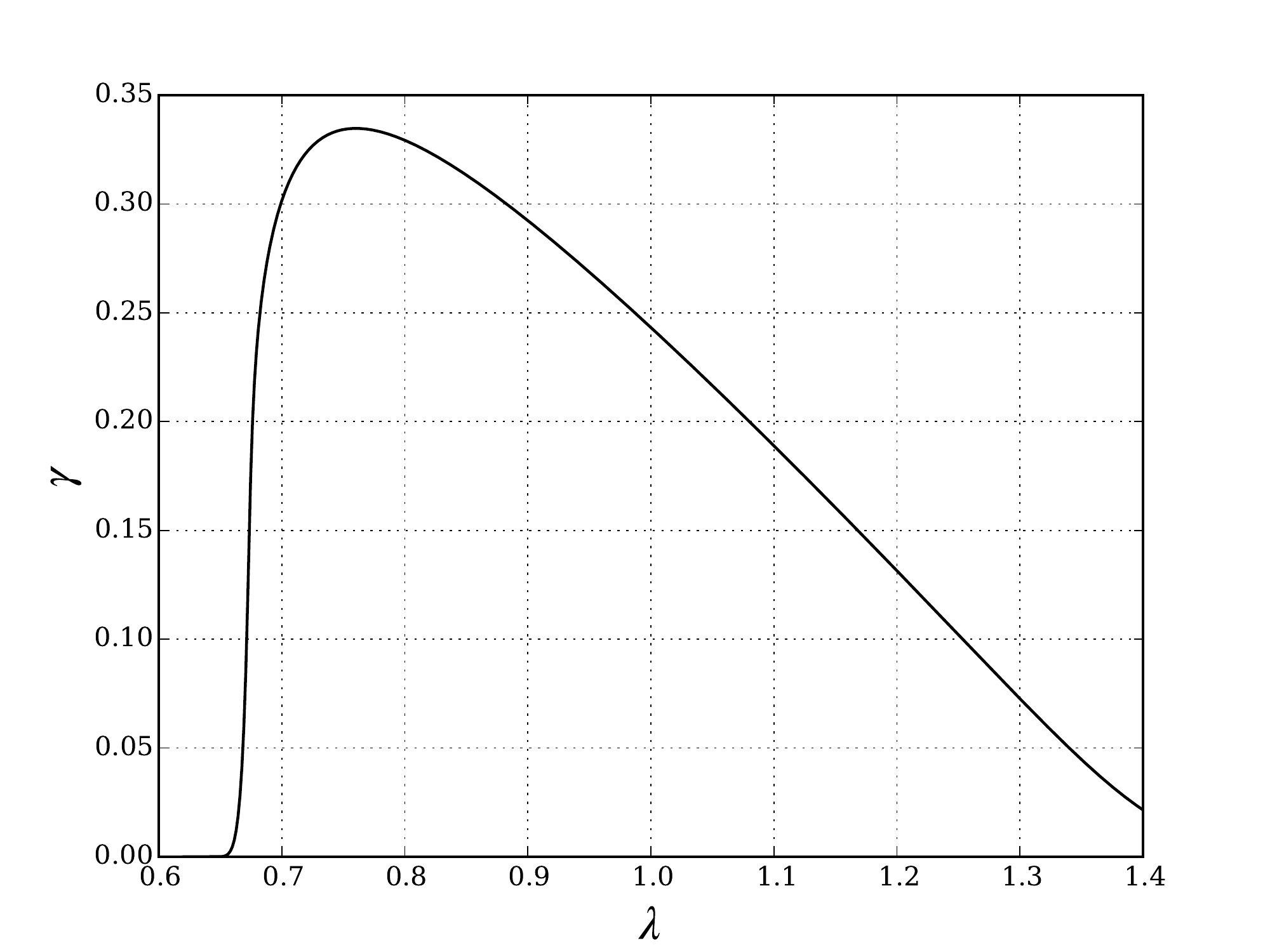}\\
        \includegraphics[width=0.70\textwidth]{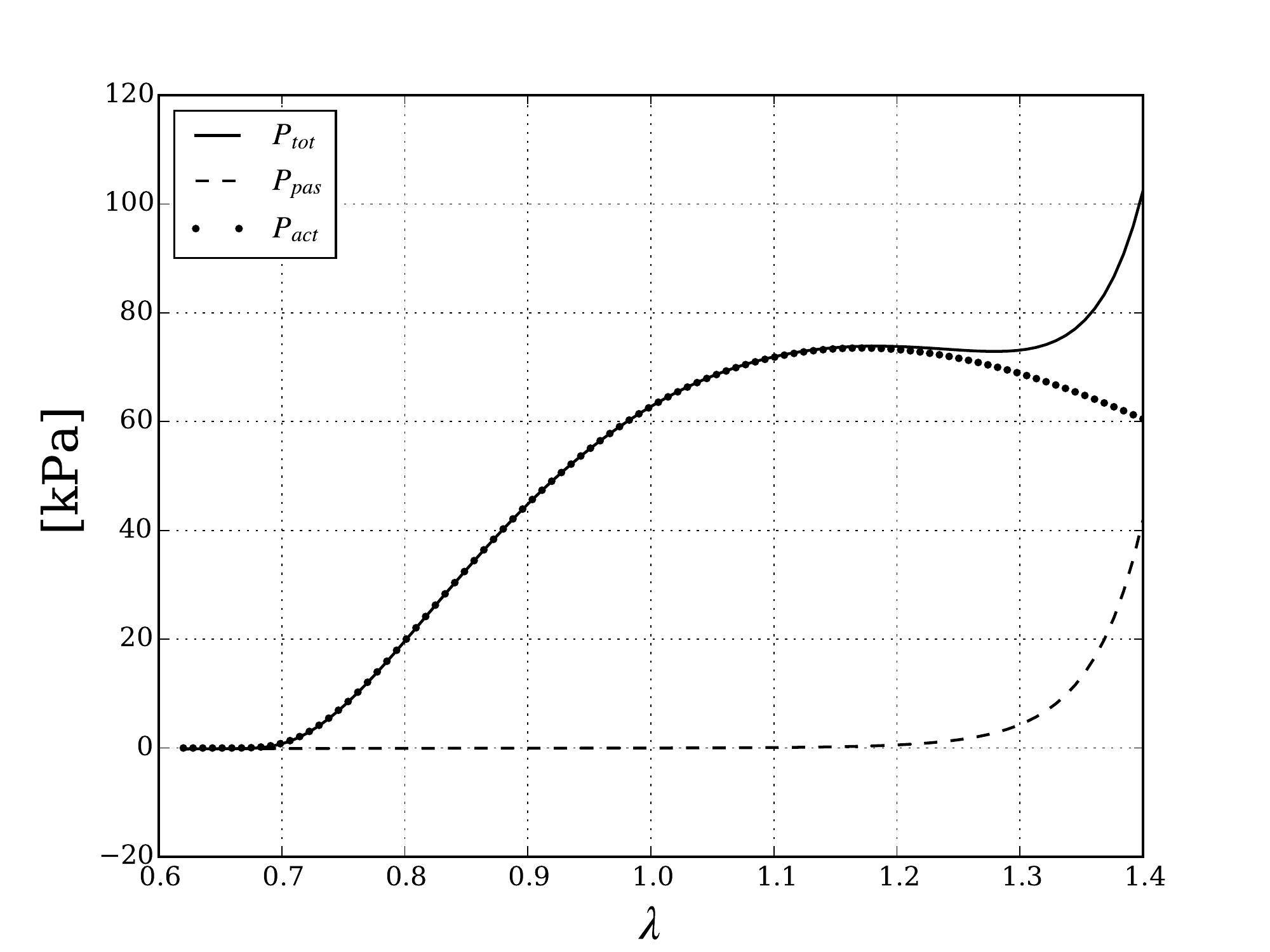}
    \end{tabular}
   \caption{The first figure shows the behavior of $\gamma$ when $\lambda$ varies: the corresponding plots of $P_{tot}$ and $P_{act}$ are given in the second figure, together with $P_{pas}$.} 
     \label{D:pactgamma}
\end{figure}

Our aim is to find the value of $\gamma$ such that
\begin{equation}
{P}_{tot}(\lambda, \gamma)= {P}_{act}(\lambda)+{P}_{pas}(\lambda), \label{D:eqgammaimplicita}
\end{equation}
where ${P}_{act} (\lambda)$ is given by \eqref{D:pact}. Unfortunately,
this leads to an equation for $\gamma$ which cannot be explicitly
solved:
\begin{eqnarray}
(1-\gamma)\left\{\left[\left(1-\frac{2}{3}w_0\right)\frac{\lambda}{(1-\gamma)^2}
-\frac{w_0}{3}\frac{1}{\lambda^2}
\right]
    e^{\alpha\left[
    \left(1-\frac{2}{3}w_0\right)\frac{\lambda^2}{(1-\gamma)^2}+\frac{w_0}{3}\frac{2}{\lambda}
    -1\right]} \nonumber \right.\\[1ex]
    \left.+\frac{w_0}{3}
-\left(1-\frac{2}{3}w_0\right)\frac{(1-\gamma)^2}{\lambda^3}\right\}
    =\frac{2}{\mu}\left[{P}_{act}(\lambda)+{P}_{pas}(\lambda)\right]. \label{D:eqgammaesplicita}
\end{eqnarray}
However one can employ standard numerical methods and plot the
solution. Fig.~\ref{D:pactgamma}$_1$, which is obtain by a bisection
method, shows $\gamma$ as a function of $\lambda$.
We remark that $\gamma$ vanishes before $\lambda_{min}$, indeed in
this region there is no difference between total and passive
stress. The corresponding behavior of the stresses is plotted in
Fig.~\ref{D:pactgamma}$_2$, which is very similar to the
representative plot of Fig.~\ref{D:attivovspassivo}.

We emphasize that the previous model is not strictly
hyperelastic, since in the expression of the
stress~\eqref{D:stressPKtotale} the derivative of $\gamma$ with
respect to $\mathbf{F}$ has been neglected. We are now working on a
truly hyperelastic model, which can be useful for some numerical
implementations.

\subsection{Loss of activation} 
\label{D:subsec:lossactivation}\index{activation!loss of}

We now want to describe from a mathematical point of view the loss of
performance of a skeletal muscle tissue. As we have already explained
in the Introduction, this is one of the main effects of
\emph{sarcopenia},\index{sarcopenia} which is a typical syndrome of
advanced age.

In \cite{D:Lang2010,D:overviewSarco} it is remarked that aging is
associated with changes in muscle mass, composition, activation and
material properties. In sarcopenic muscle, there is a loss of motor
units \emph{via} denervation and a net conversion in slow fibers, with a
resulting loss in muscle power. Hence, the loss of performance of a
sarcopenic muscle can be described as a weakening of the activation of
the fibers.

\begin{figure}[htbp]
\centering
        \includegraphics[width=0.70\textwidth]{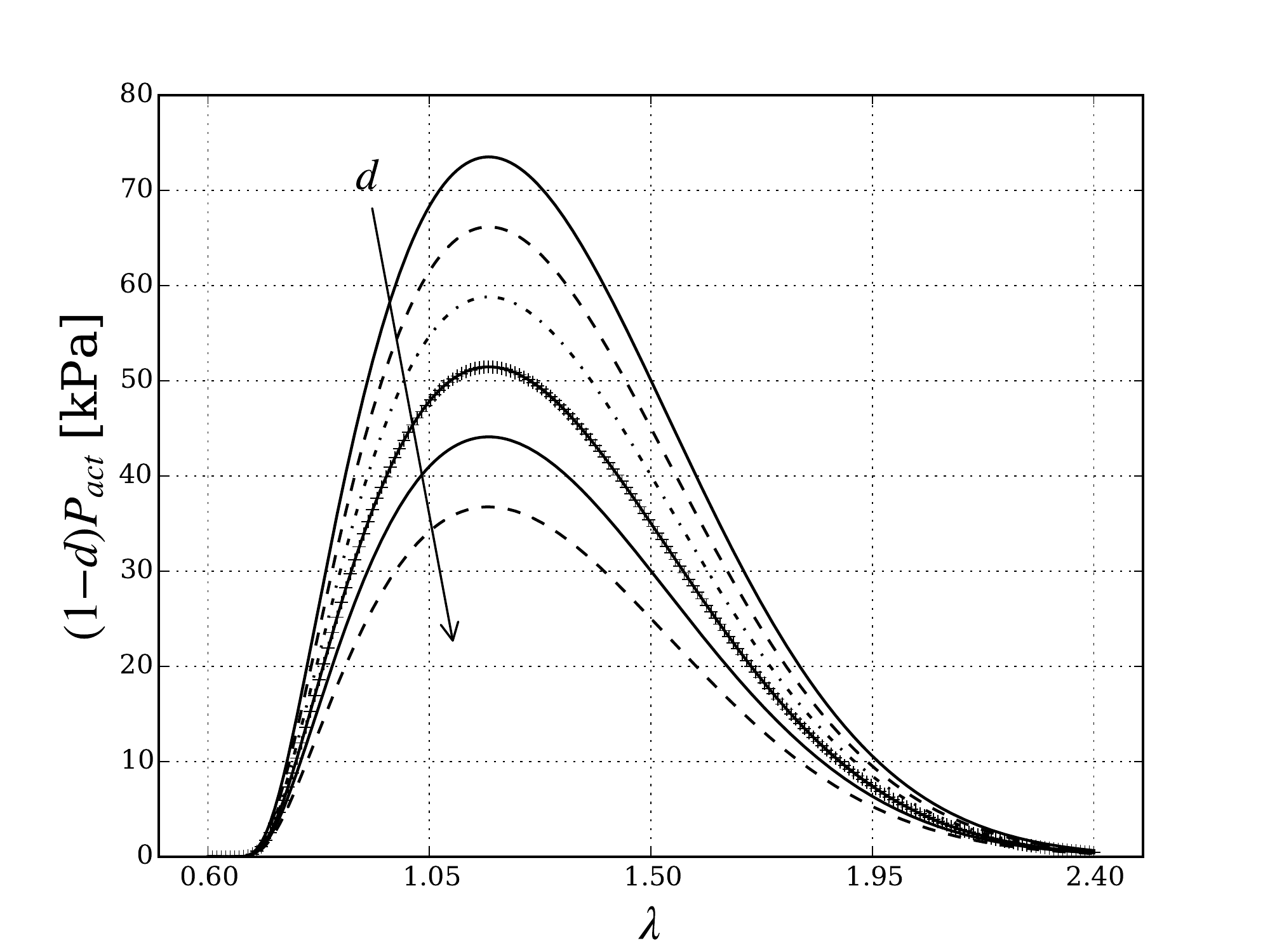}
   \caption{Plot of $(1-d)P_{act}(\lambda)$ vs $\lambda$ when $d$ varies from $0$ to $0.5$.} 
     \label{D:figdpact}
\end{figure}

Unfortunately, as far as we know, there are no experimental data
describing a uniaxial simple tension along the fibers of a sarcopenic
muscle. For this reason, we try to describe the loss of activation by
a parameter $d$ which lowers the curve $P_{act}(\lambda)$ given by
\eqref{D:pact}. The parameter $d$ describes the percentage of
\emph{disease} or \emph{damage}: if $d=0$, then the muscle is
healthy. In order to get our aim, we multiply the function
$P_{act}(\lambda)$ by the factor $1-d$, as one can see in
Fig.~\ref{D:figdpact}. Notice that such a choice can be overly simple:
for instance, it implies that the maximum is always attained at
$\lambda_{opt}$, even if there is no experimental evidence of
that. However, the presence of $d$ allows to describe, at least
qualitatively, the loss of performance of a muscle, which is one of the
goals of our model.

\section{Numerical validation}
\label{D:sec:numerical}

Finally, we simulate numerically the contraction and the elongation of
a slab of skeletal muscle tissue represented by a cylinder. We assume
radial symmetry, so that the mesh is a rectangle.  The ends of the
cylinder are assumed to remain perpendicular to the axial direction.
The rectangle is modeled by the hyperelastic model presented in the
previous sections. The active contractile fibers are aligned along the
length of the rectangle, which coincides with ${\e}_1$.  The passive
and active material parameters are given in Tables \ref{D:parpass} and
\ref{D:paratt}, respectively. Concerning the boundary conditions, the
cylinder is fixed at one end and elongated to a given length, in order
to recreate the situation of the experiments reported in
\cite{D:datib}. The lateral surface is assumed to be tension-free.

The analysis is performed by using the computing environment
FEniCS. The FEniCS Project\index{FEniCS} \cite{D:fenics} is a collection of
numerical software, supported by a set of novel algorithms and
techniques, aimed at the automated solution of differential equations
using finite element methods.

As it is explained in Section \ref{D:subsec:gammavselongation}, one of
the main features of our model is the dependence of the activation
parameter $\gamma$ on the stretch $\lambda$. The function
$\gamma(\lambda)$ solves the implicit equation
\eqref{D:eqgammaesplicita}, which ensures that the corresponding
stress curves fit the experimental data. However, even if this
equation can be solved using numerical methods, it is interesting to
find an explicit function in order to analyze qualitatively the active
model and to run the simulations in FEniCS. Moreover, the explicit
function $\gamma(\lambda)$ has to be very precise, since a slight
error on $\gamma$ deeply affects the behavior of the total stress.
Hence, it is reasonable to relate the expression of $\gamma$ to 
the material parameters and the quantities involved in
\eqref{D:eqgammaesplicita}.
An idea is to isolate the exponential in \eqref{D:eqgammaesplicita} 
and to express its exponent by a first step approximation of a
fixed-point method. We then obtain the following expression of
$\gamma$:
\begin{eqnarray}
&\gamma(\lambda)=&
\left\{
\begin{aligned}
&a\left[\sqrt{\frac{1-\frac{2}{3}w_0}{\frac{g(\lambda_{min})}{\alpha}+\frac{w_0}{3\lambda_{min}}}}\lambda_{min}
-\sqrt{\frac{1-\frac{2}{3}w_0}{\frac{g(\lambda)}{\alpha}+\frac{w_0}{3\lambda}}}\lambda
\right]
& \text{if $\lambda>\lambda_{min}$},
\\[1ex]
&0 & \text{otherwise},
\end{aligned}
\right.
 \label{D:gamma}\\[1ex]
&g(\lambda)=&\ln\alpha+\alpha\left(1-\frac{w_0}{\lambda}\right)-\frac12\ln\left(\frac{1-\frac{2}{3}w_0}{\frac{1}{\alpha}+\frac{w_0}{3\lambda}}\right)\nonumber\\[1ex]
&&+\ln\left\{b\frac{2}{\mu}\left[P_{act}(\lambda)+P_{pas}(\lambda)\right]+
\sqrt{\frac{1-\frac{2}{3}w_0}{\frac{1}{\alpha}+\frac{w_0}{3\lambda}}}\left[
\frac{\left(1-\frac{2}{3}w_0\right)^2}{\frac{1}{\alpha}+\frac{w_0}{3\lambda}}-\lambda\frac{w_0}{3} 
\right]\right\}, \nonumber
\end{eqnarray}
where $a$ and $b$ are dimensionless fitting parameters: $a$ is related to the magnitude
of $\gamma$, while $b$ acts on the curves \eqref{D:pact} and
\eqref{D:Ppas}, which are the terms of the equation not depending on
$\gamma$.  Performing a least square optimization on the resulting
$P_{act}$, one gets $a=1.0133$ and $b=0.2050$.

\begin{figure}[htbp]
        \begin{tabular}{cc}
        \includegraphics[width=0.70\textwidth]{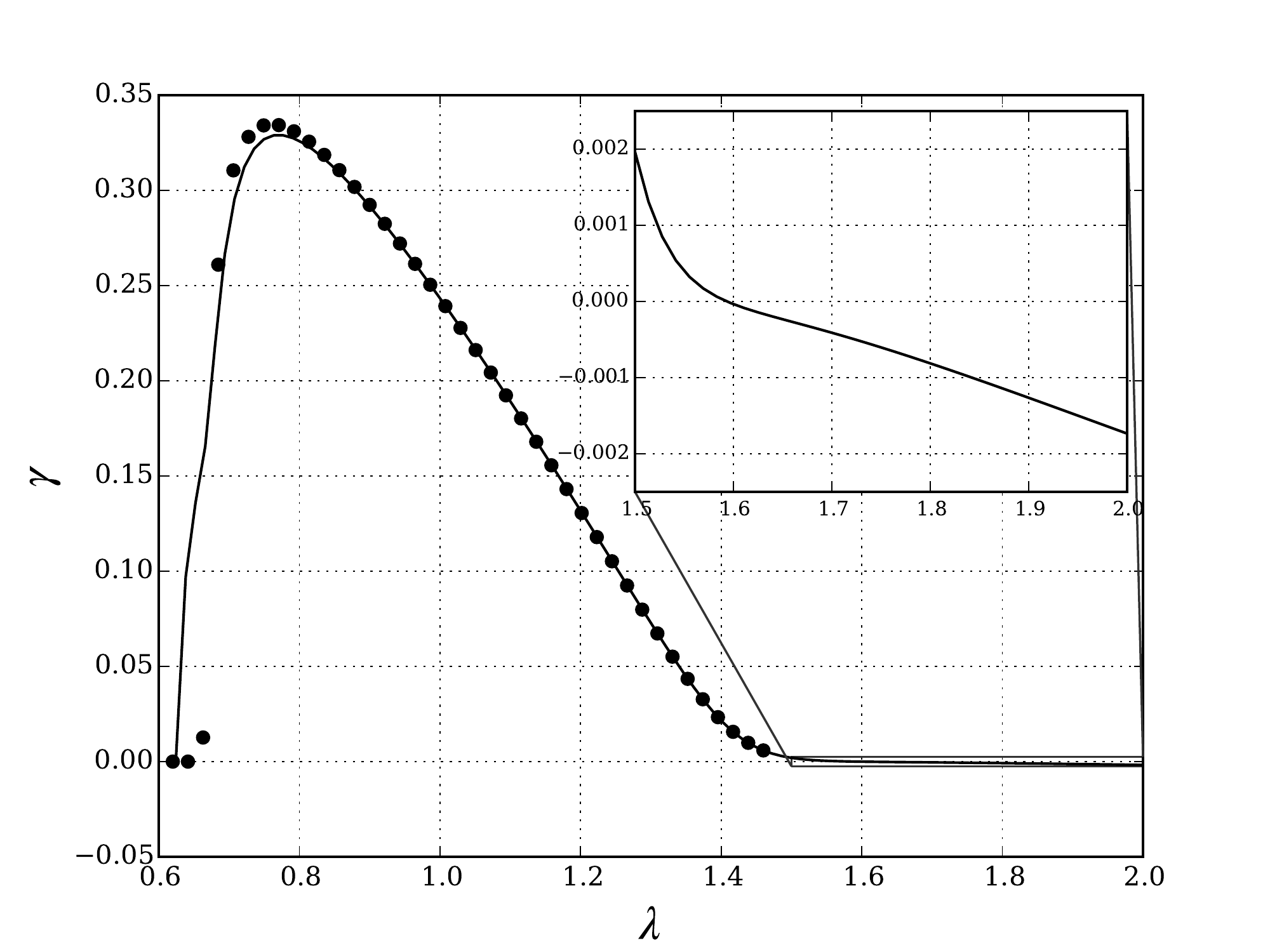}
    \end{tabular}
   \caption{Comparison between the behavior of $\gamma(\lambda)$ in
     \eqref{D:gamma} (solid line) and the numerical solution of equation \eqref{D:eqgammaesplicita} (dotted line).} 
     \label{D:figapprgamma}
\end{figure}

\begin{figure}[htbp]
        \begin{tabular}{cc}
        \includegraphics[width=0.70\textwidth]{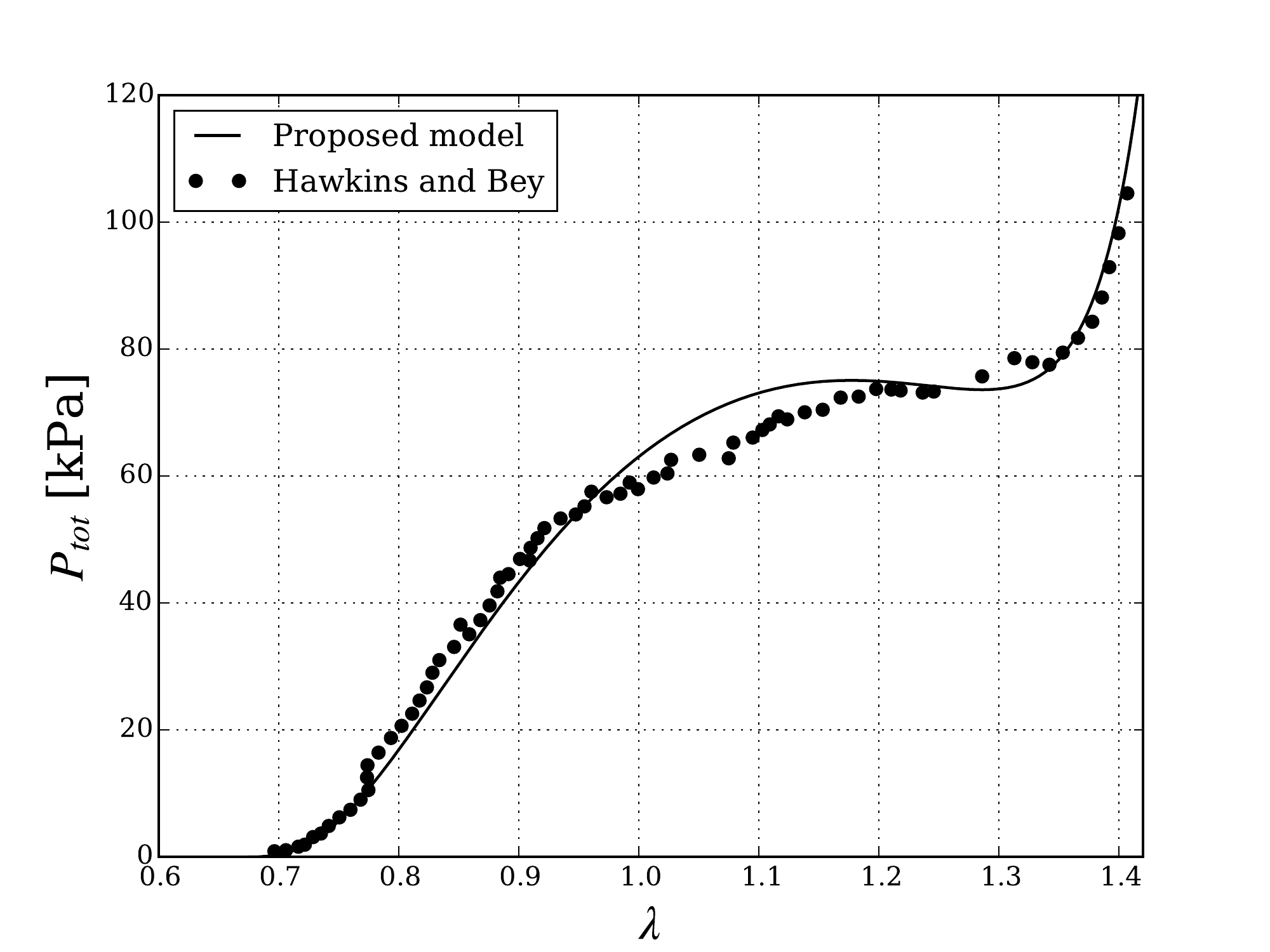}
    \end{tabular}
   \caption{Trend of $P_{tot}$ when $\gamma$ is given by \eqref{D:gamma} and $\lambda$ varies.} 
     \label{D:figapprgammaptot}
\end{figure}

Fig.~\ref{D:figapprgamma} shows the plot of the function
$\gamma(\lambda)$ given in \eqref{D:gamma} in comparison to the
numerical solution of equation \eqref{D:eqgammaesplicita} obtained by
a bisection method.
Notice that the function defined in~\eqref{D:gamma} is continuous; in
particular we impose $\gamma(\lambda_{min})=0$, so that the starting
value of activation does not change. Moreover, the function
approximates very well the numerical values of $\gamma$ in the range
$0.7<\lambda<1.5$. However, the fitting is not so good when $\lambda$
becomes larger: for instance, the function is negative for
$\lambda\geq 1.6$. Nevertheless, the latter behavior of $\gamma$ does not
influence too much the curve $P_{tot}$, since in that region
$P_{pas}\gg P_{act}$. Indeed, one can even neglect the activation for
large stretches.
The total stress response  is plotted in Fig.~\ref{D:figapprgammaptot}
in comparison to the data given in \cite{D:datib}.

\begin{figure}[htbp]
        \begin{tabular}{cc}
        \includegraphics[width=0.70\textwidth]{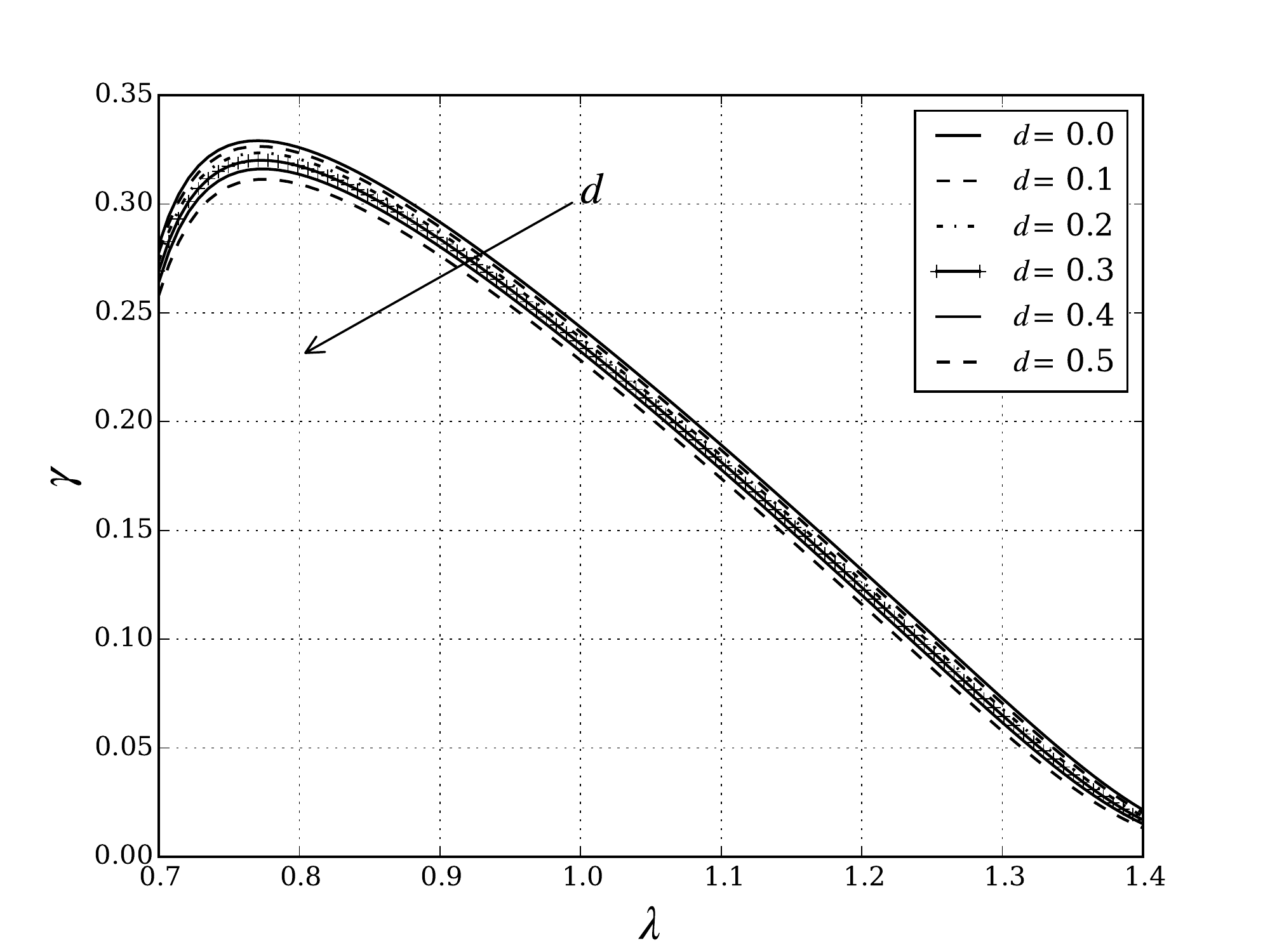}\\
        \includegraphics[width=0.70\textwidth]{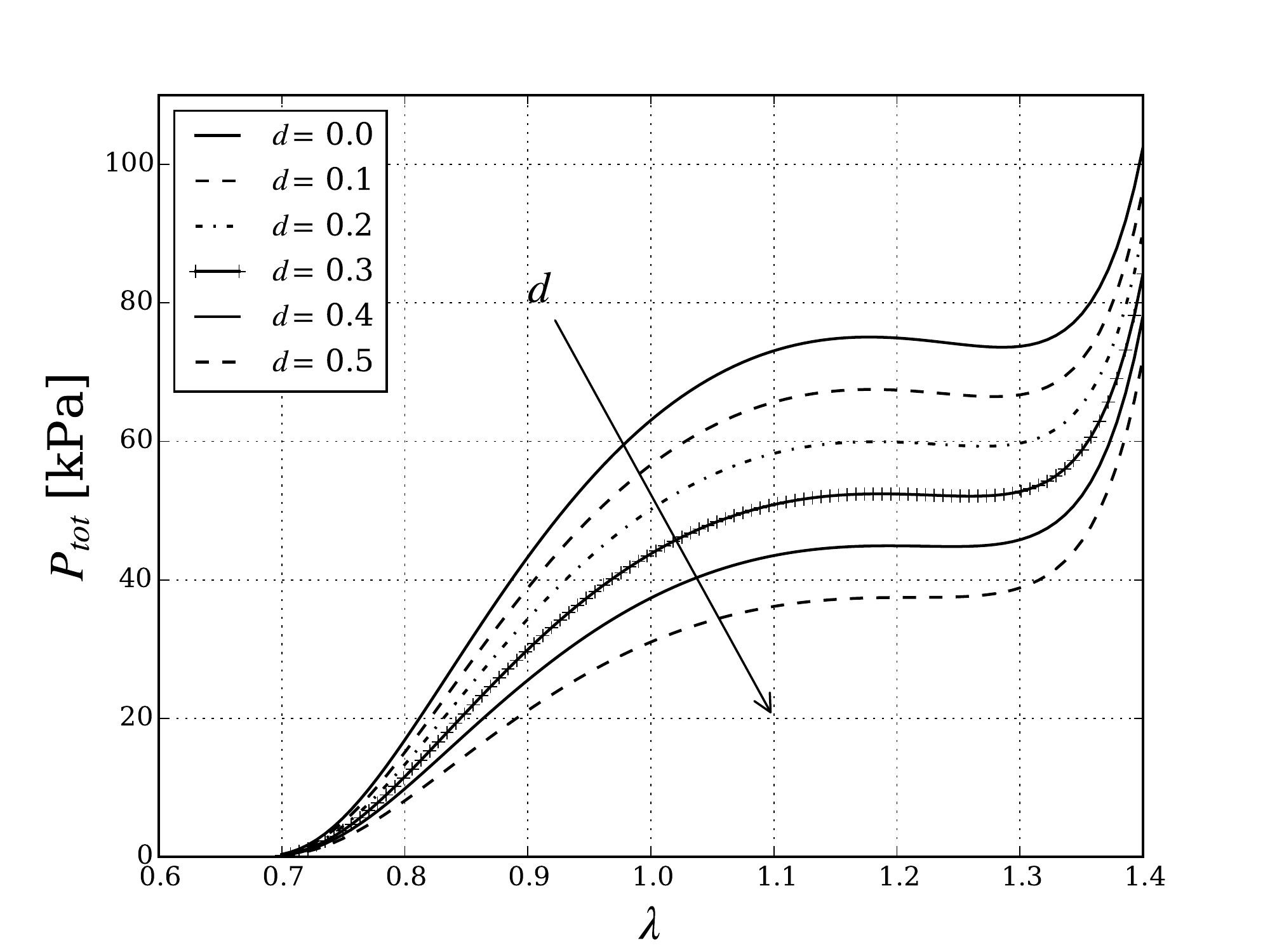}
    \end{tabular}
   \caption{Behavior of $\gamma$ and $P_{tot}$ when $\lambda$ varies.} 
     \label{D:figfenics}
\end{figure}

Finally, it is interesting to run the simulations in the case of loss
of activation, \emph{i.e.} when the damage parameter $d$ varies. In
order to find the suitable activation function $\gamma(\lambda)$, it
is sufficient to multiply the term $P_{act}$ in~\eqref{D:gamma} by
$(1-d)$.  As one would expect from Fig.~\ref{D:figdpact}, we have that
when $d$ increases the activation $\gamma$ decreases
(Fig.~\ref{D:figfenics}$_1$). This means that lowering the curve of
$P_{act}$ results in a decrease of $\gamma(\lambda)$, which leads to a
lowered total stress response. As one can see in
Fig.~\ref{D:figfenics}$_2$, the damage parameter mainly affects the
value of the stress in the region near $\lambda_{opt}$, where the
active stress reaches its maximum. However, the qualitative behavior of the stress curve
does not change, at least for $d\leq 0.5$. In particular, after a
plateau, the stress follows the exponential growth of the passive curve.

\section*{Acknowledgement}
  This work has been supported by the project \emph{Active Ageing and
    Healthy Living} \cite{D:progettob} of the Universit\`a Cattolica
  del Sacro Cuore and partially supported by GNFM (Gruppo
  Nazionale per la Fisica Matematica) of INdAM (Istituto Nazionale di
  Alta Matematica).

The authors wish to thank the anonymous referees for their useful comments.


\begin{thebibliography}{10}

\bibitem{D:fenics}
M.~S. Aln\ae{}s, J.~Blechta, J.~Hake, A.~Johansson, B.~Kehlet, A.~Logg,
  C.~Richardson, J.~Ring, M.~E. Rognes, and G.~N. Wells.
\newblock The {FE}ni{CS} {P}roject {V}ersion 1.5.
\newblock {\em Archive of Numerical Software}, 100:9--23, 2015.

\bibitem{D:asasb}
D.~Ambrosi and S.~Pezzuto.
\newblock {A}ctive {S}tress vs. {A}ctive {S}train in {M}echanobiology:
  {C}onstitutive {I}ssues.
\newblock {\em Journal of Elasticity}, 107:199--212, 2012.

\bibitem{D:blemker}
S.~S. Blemker, P.~M. Pinsky, and S.~L. Delp.
\newblock A 3{D} model of muscle reveals the causes of nonuniform strains in
  the biceps brachii.
\newblock {\em Journal of Biomechanics}, 38:657--665, 2005.

\bibitem{D:bol}
M.~B\"{o}l and S.~Reese.
\newblock Micromechanical modelling of skeletal muscles based on the finite
  element method.
\newblock {\em Computer Methods in Biomechanics and Biomedical Engineering},
  11:489--504, 2008.

\bibitem{D:review}
G.~Chagnon, M.~Rebouah, and D.~Favier.
\newblock {H}yperelastic {E}nergy {D}ensities for {S}oft {B}iological
  {T}issues: {A} {R}eview.
\newblock {\em Journal of Elasticity}, 120:129--160, 2015.

\bibitem{D:reportSarcopenia}
A.~J. Cruz-Jentoft, J.~P. Baeyens, J.~M. Bauer, Y.~Boirie, T.~Cederholm,
  F.~Landi, F.~C. Martin, J.~P. Michel, Y.~Rolland, S.~M. Schneider,
  E.~Topinkov\'{a}, M.~Vandewoude, and M.~Zamboni.
\newblock Sarcopenia: European consensus on definition and diagnosis.
\newblock {\em Age and Ageing}, 39:412--423, 2010.

\bibitem{D:ebi}
A.~E. Ehret, M.~B\"{o}l, and M.~Itskov.
\newblock A continuum constitutive model for the active behaviour of skeletal
  muscle.
\newblock {\em Journal of the Mechanics and Physics of Solids}, 59:625--636,
  2011.

\bibitem{D:ebipcb}
A.~E. Ehret and M.~Itskov.
\newblock A polyconvex hyperelastic model for fiber-reinforced materials in
  application to soft tissues.
\newblock {\em Journal of Materials Science}, 42:8853--8863, 2007.

\bibitem{D:ei2009}
A.~E. Ehret and M.~Itskov.
\newblock Modeling of anisotropic softening phenomena: {A}pplication to soft
  biological tissues.
\newblock {\em International Journal of Plasticity}, 25:901--919, 2009.

\bibitem{D:datib}
D.~Hawkins and M.~Bey.
\newblock {A} {C}omprehensive {A}pproach for {S}tudying {M}uscle-{T}endon
  {M}echanics.
\newblock {\em ASME Journal of Biomechanical Engineering}, 116:51--55, 1994.

\bibitem{D:thomas}
T.~Heidlauf and O.~R\"{o}hrle.
\newblock A multiscale chemo-electro-mechanical skeletal muscle model to
  analyze muscle contraction and force generation for different muscle fiber
  arrangements.
\newblock {\em Frontiers in Physiology}, 5:1--14, 2014.

\bibitem{D:hernandez}
B.~Hern\'andez-Gasc\'on, J.~Grasa, B.~Calvo, and J.~F. Rodr\'iguez.
\newblock A 3{D} electro-mechanical continuum model for simulating skeletal
  muscle contraction.
\newblock {\em Journal of Theoretical Biology}, 335:108--118, 2013.

\bibitem{D:Johansson}
T.~Johansson, P.~Meier, and R.~Blickhan.
\newblock {A} {F}inite-{E}lement {M}odel for the {M}echanical {A}nalysis of
  {S}keletal {M}uscles.
\newblock {\em Journal of Theoretical Biology}, 206:131--149, 2000.

\bibitem{D:Lang2010}
T.~Lang, T.~Streeper, P.~Cawthon, K.~Baldwin, D.~R. Taaffe, and T.~B. Harris.
\newblock Sarcopenia: etiology, clinical consequences, intervention, and
  assessment.
\newblock {\em Osteoporos Int}, 21:543--559, 2010.

\bibitem{D:martins}
J.~A.~C. Martins, E.~B. Pires, R.~Salvado, and P.~B. Dinis.
\newblock A numerical model of passive and active behavior of skeletal muscles.
\newblock {\em Computer Methods in Applied Mechanics and Engineering},
  151:419--433, 1998.

\bibitem{D:giulio}
A.~Musesti, G.~G. Giusteri, and A.~Marzocchi.
\newblock {P}redicting {A}geing: {O}n the {M}athematical {M}odelization of
  {A}geing {M}uscle {T}issue.
\newblock In G.~Riva et~al., editor, {\em Active Ageing and Healthy Living}.
  IOS press, 2014.
\newblock Chapter 17.

\bibitem{D:nardinocchiteresi}
P.~Nardinocchi and L.~Teresi.
\newblock {O}n the {A}ctive {R}esponse of {S}oft {L}iving {T}issues.
\newblock {\em Journal of Elasticity}, 88:27--39, 2007.

\bibitem{D:PTD}
C.~Paetsch, B.~A. Trimmer, and A.~Dorfmann.
\newblock A constitutive model for active-passive transition of muscle fibers.
\newblock {\em International Journal of Non-Linear Mechanics}, 47:377--387,
  2012.

\bibitem{D:progettob}
G.~Riva, P.~Ajmone~Marsan, and C.~Grassi.
\newblock {\em {A}ctive {A}geing and {H}ealthy {L}iving}.
\newblock IOS press, 2014.

\bibitem{D:sch}
J.~Schr\"{o}der and P.~Neff.
\newblock Invariant formulation of hyperelastic transverse isotropy based on
  polyconvex free energy functions.
\newblock {\em International Journal of Solids and Structures}, 40:401--445,
  2003.

\bibitem{D:taber}
L.~A. Taber and R.~Perucchio.
\newblock {M}odeling {H}eart {D}evelopment.
\newblock {\em Journal of Elasticity}, 61:165--197, 2000.

\bibitem{D:vanLeeuwen1991}
J.~L. van Leeuwen.
\newblock Optimum power output and structural design of sarcomeres.
\newblock {\em Journal of Theoretical Biology}, 149:229--256, 1991.

\bibitem{D:vanLeeuwen1992}
J.~L. van Leeuwen.
\newblock Muscle function in locomotion.
\newblock In {\em Advances in Comparative and Environmental Physiology}.
  Springer Heidelberg Berlin, 1992.
\newblock Chapter 7.

\bibitem{D:overviewSarco}
S.~von Haehling, J.~E. Morley, and S.~D. Anker.
\newblock An overview of sarcopenia: facts and numbers on prevalence and
  clinical impact.
\newblock {\em Journal of Cachexia, Sarcopenia and Muscle}, 1:129--133, 2010.

\end{thebibliography}
\end{document}